\documentclass[prd,preprint,tightenlines,floatfix,showpacs,preprintnumbers,nofootinbib,eqsecnum]{revtex4}
\usepackage[dvips,final]{graphicx}
\usepackage{epsfig}
\usepackage{bm}

\begin{document}

\date{\today}
\title{Power corrections to the $\pi^0 \gamma$ transition
form factor and pion distribution amplitudes}
\author{\textbf{S.~S.~Agaev}}
\email{agaev_shahin@yahoo.com}
\affiliation { High Energy Physics Lab.,
               Baku State University,
	        Z. Khalilov st. 23, 370148 Baku, Azerbaijan}

\begin{abstract}
Employing the standard hard-scattering approach and the running
coupling method we calculate a class of power-suppressed corrections 
$\sim 1/Q^{2n},\,n=1,2,3\ldots$ to the electromagnetic $\pi ^0\gamma $
transition form factor (FF) $Q^2F_{\pi \gamma }(Q^2)$ arising from the
end-point $x \to 0,\;1$ integration regions. In the investigation we
use a hard-scattering amplitude of the subprocess 
$\gamma +\gamma^{*}\rightarrow q+\overline{q}$, 
symmetrized under the exchange $\mu_R^2\leftrightarrow \overline{\mu }_R^2$ 
important for exclusive processes containing two external photons. 
In the computations the pion model distribution amplitudes (DA's) with one 
and two nonasymptotic terms are employed. The obtained predictions are compared 
with the CLEO data and constraints on the DA parameters $b_2(\mu _0^2)$ 
and $b_4(\mu _0^2)$ at the normalization point $\mu _0^2=1\,{\rm GeV}^2$ 
are extracted. Further restrictions on the pion DA's are deduced from 
the experimental data on the electromagnetic FF $F_\pi (Q^2)$.\\

\end{abstract}


\maketitle

\newpage

\section{Introduction}
\setcounter{equation}0

The $\pi ^0$ meson electromagnetic transition form factor (FF) $F_{\pi
\gamma }(Q^2)$ is among the simplest exclusive processes for investigation
of which at large momentum transfer the perturbative QCD (PQCD) methods \cite
{br1,rad1,dun} can be applied. Because of the recent CLEO data \cite{cleo},
where the form factor $F_{\pi \gamma }(Q^2)$ was measured with high
precision, the interest in this process has been renewed. Thus during the last
few years for computation of $F_{\pi \gamma }(Q^2)$ the various theoretical
methods and schemes were proposed \cite{ag1,stef2,yak,eks,ag2}. The
aim here is twofold: to elaborate methods for the calculation of 
$F_{\pi \gamma}(Q^2)$ within PQCD and, at the same time, to extract from experimental data
information on the pion distribution amplitude (DA). The latter, being 
independent of a specific exclusive process and universal quantity, is
an important input ingredient in studying various processes that involve the
pion.

It is known that in experiments the $\pi ^0\gamma $ transition was explored
at momentum transfers of $Q^2\sim 1-10\,{\rm GeV}^2$, which are far from
the asymptotic limit $Q^2\rightarrow \infty $, where the PQCD\
factorization formula with the pion leading-twist asymptotic DA leads to
reliable predictions. In the present experimentally accessible energy
regimes, power-suppressed corrections $\sim 1/Q^{2n},\,n=1,2,..$ play an
important role in explaining the experimental data \cite{ag1,ag2}. There
are numerous sources of power corrections to $F_{\pi \gamma }(Q^2)$. For
example, the pion higher-twist (HT) DA's and higher Fock states generate
such corrections. Power corrections can also originate from the end-point
regions $x\rightarrow 0,\,1$ as a result of the integration of the PQCD
factorization expression with the QCD running coupling $\alpha _{{\rm s}%
}(Q^2x)$ [$\alpha _{{\rm s}}(Q^2\overline{x}),\,\overline{x}=1-x$] over the pion's quark
longitudinal momentum fraction $x$. In fact, in order to reduce the 
higher-order corrections to a physical quantity and improve the convergence of the
corresponding perturbation series, the renormalization scale $\mu _R^2$
($\overline{\mu }_R^2$), i.e., the argument of the QCD coupling, in a 
Feynman diagram  should be set equal to the virtual parton's squared
four-momentum \cite{br2}. In the exclusive
processes the scale $\mu _R^2$ ($\overline{\mu }_R^2$)
chosen this way inevitably depends on the longitudinal momentum fractions
carried by the
hadron constituents. For the photon-meson transition we have $\mu _R^2=Q^2x$
and $\overline{\mu }_R^2=Q^2\overline{x}$, because at two leading order
diagrams of the partonic subprocess $\gamma +\gamma ^{*}\rightarrow
q+\overline{q}$, absolute values of the virtual quark and antiquark
squared four-momenta are
determined by these expressions. But then the PQCD factorization formula
diverges, since $\alpha _{{\rm s}}(Q^2x)$ [$\alpha _{{\rm s}}(Q^2\overline{x}
)$] suffers from end-point $x\to  0$ [$x \to 1$] singularity. The running
coupling (RC) method
solves this problem by using a Borel transformation and applying the
principal value prescription. As a result, one obtains the Borel resummed
expression for the $\pi ^0\gamma $ transition FF, which contains 
power-suppressed corrections. The RC method in conjunction with the infrared (IR)
renormalon calculus was used for computation of such power corrections to
the $\pi ^0\gamma $ and $\eta \gamma ,\,\eta ^{\prime }\gamma $ transition
FF's \cite{ag1,ag2}, to the electromagnetic FF's of the light mesons $%
F_M(Q^2)$ ($M=\pi ,\,K,\,\rho _L$) \cite{ag3,ag4,ag7,stef3}, as well as to
the gluon-gluon-$\eta ^{\prime }$ meson vertex function \cite{ag5}.

In the present work we compute power corrections to the $\pi ^0\gamma $
transition FF employing the version of the hard-scattering amplitude 
symmetrized under replacement $\mu_R^2\leftrightarrow \overline{\mu }_R^2$. 
The symmetrization procedure is important for
exclusive processes with two external photons (gluons) in the
hard-scattering Feynman diagrams, because it allows one to treat within the
RC method both virtual and real photons (gluons) on the same footing. The
latter is required in order to consider the $\pi ^0\gamma ^{*}$ and $\pi
^0\gamma $ transitions in a unifying way, i.e., to get in the limits $\omega
\rightarrow 0;1$ ($\omega $ is the asymmetry parameter) from the $\pi
^0\gamma ^{*}$ transition FF $F_{\pi \gamma ^{*}}(Q^2,\omega )$ the FF $%
F_{\pi \gamma }(Q^2)$ of the $\pi ^0\gamma $ transition. The advocated
method was used in our previous work \cite{ag5} to investigate the virtual
and on-shell gluon-$\eta ^{\prime }$ meson transitions. In what follows we
refer to this approach as the symmetrized RC (SRC) method.

This paper is organized as follows: in Sec. II we introduce the
symmetrization procedure of the hard-scattering amplitude for the $\pi^0\gamma $ 
transition. Here we present our results for the Borel resummed 
$[Q^2F_{\pi \gamma }(Q^2)]^{res}$ FF obtained within the SRC
method. Section III is devoted to detailed analysis of the $Q^2\rightarrow
\infty $ limit of $[Q^2F_{\pi \gamma }(Q^2)]^{res}$. In Sec. IV we perform
numerical computations and from comparison of our predictions with the CLEO
data extract constraints on the parameters $b_2^0(1\,{\rm GeV}^2)$ and $%
b_2^0(1\,{\rm GeV}^2),\,\,\,b_4^0(1\,{\rm GeV}^2)$ in the pion DA's with one
and two nonasymptotic terms, respectively. Further restrictions
on the DA's arising from analysis of the pion electromagnetic FF 
are described in Sec. V. In Sec. VI we make our
concluding remarks.

\section{The photon-meson transition form factor}

\setcounter{equation}0

\subsection{The symmetrized version of the hard-scattering amplitude}

The real photon-pseudoscalar $M$ meson electromagnetic transition FF $%
F_{M\gamma }(Q^2)$ can be defined in terms of the amplitude $\Gamma ^{\mu
\nu }$, 
\begin{equation}
\label{eq:2.1}\Gamma ^{\mu \nu }=ie^2F_{M\gamma }(Q^2)\epsilon ^{\mu \nu
\alpha \beta }P_\alpha q_{1\beta }, 
\end{equation}
for the process 
\begin{equation}
\label{eq:2.2}\gamma ^{*}(q_1)+\gamma (q_2)\rightarrow M(P), 
\end{equation}
where $Q^2=-q_1^2$ is the momentum transfer.

At the large momentum transfer the FF $F_{M\gamma }(Q^2)$ is given by the
factorization formula of the standard hard-scattering approach (HSA) \cite
{br1}, 
\begin{equation}
\label{eq:2.3}F_{M\gamma }(Q^2)=\left[ T_H^1\left( x,Q^2,\mu _F^2\right)
+T_H^2\left( x,Q^2,\mu _F^2\right) \right] \otimes \phi _M(x,\mu _F^2). 
\end{equation}
Here the function $T_H(x,Q^2,\mu _F^2)$, 
\begin{equation}
\label{eq:2.4}T_H\left( x,Q^2,\mu _F^2\right) =T_H^1\left( x,Q^2,\mu
_F^2\right) +T_H^2\left( x,Q^2,\mu _F^2\right) , 
\end{equation}
is the hard-scattering amplitude of the subprocess $\gamma +\gamma
^{*}\rightarrow q+\overline{q}$ , $\phi _M(x,\mu _F^2)$ is the meson DA, $%
\mu _F^2$ is the factorization scale, and $\overline{x}\equiv 1-x$, $x$ being
the longitudinal momentum fraction carrying by the meson's quark. In Eq.\ (%
\ref{eq:2.3}) the shorthand notation 
\begin{equation}
\label{eq:2.5}T_H\left( x,Q^2,\mu _F^2\right) \otimes \phi _M(x,\mu
_F^2)=\int_0^1T_H\left( x,Q^2,\mu _F^2\right) \phi _M(x,\mu _F^2)dx 
\end{equation}
is used.

It is evident that a physical quantity, represented by the factorization
formula, Eq.\ (\ref{eq:2.3}) being a sample one, does not depend on renormalization 
and factorization schemes and scales employed for its calculation.
But at any finite order of the QCD perturbation theory, due to truncation of
the corresponding perturbation series, the hard-scattering amplitude (\ref
{eq:2.4}) depends on both the factorization $\mu _F^2$ and renormalization $%
\mu _R^2$ scales. Since higher-order corrections in PQCD computations, as a
rule, are large for both inclusive and exclusive processes, in order to get
reliable theoretical predictions within the PQCD by means of the truncated
perturbation series, an optimal choice for these scales, i.e., a choice
that minimizes higher-order corrections, is required. The factorization scale 
$\mu _F^2$ in exclusive processes is traditionally set equal to the momentum
transfer $Q^2$, because higher-order corrections contain terms $\sim \ln
(\mu _F^2/Q^2)$ and such choice eliminates them in hard-scattering
amplitudes. As a result, in the factorization formula only a hadron DA
explicitly depends on the scale $\mu _F^2=Q^2$.

The situation with the renormalization scale $\mu _R^2$ is more subtle.
Really, this scale appears, in general, not only in higher-order corrections
to the hard-scattering amplitude, but also determines the scale of the QCD
coupling $\alpha _{{\rm s}}(\mu _R^2)$. In order to reduce higher-order
corrections to a physical quantity, in exclusive proceses the scale $\mu
_R^2 $ should be taken equal to the square of the momentum transfer carried
by a virtual parton in each leading order Feynman diagram of the underlying
hard-scattering subprocess \cite{br2}. For the real photon-meson transition
these scales are determined by the leading order diagrams of the subprocess $%
\gamma +\gamma ^{*}\rightarrow q+\overline{q}$ and are given by the
expressions 
\begin{equation}
\label{eq:2.5a}\mu _R^2=Q^2x,\;\;\overline{\mu }_R^2=Q^2\overline{x}. 
\end{equation}

After these remarks let us turn to our formulas (\ref{eq:2.3}) and (\ref
{eq:2.4}). In accordance with the ''tradition,'' in this work we set $\mu
_F^2=Q^2$ and in what follows omit the dependence of the hard-scattering
amplitude on the scale $\mu _F^2$. Then, for the hard-scattering amplitude
at the next-to-leading order (NLO) we get \cite{aguila}

\begin{equation}
\label{eq:2.6}T_H^1(x,Q^2,\mu _R^2)=\frac N{Q^2}\frac 1x\left[ 1+C_F\frac{%
\alpha _{{\rm s}}(\mu _R^2)}{4\pi }t(x)\right] , 
\end{equation}
where the function $t(x)$ is given by the expression 
\begin{equation}
\label{eq:2.7}t(x)=\ln {}^2x-\frac{x\ln x}{\overline{x}}-9. 
\end{equation}
Here $N$ is the constant, which depends on the quark structure of the meson, 
$C_F=4/3$ is the color factor. The second function $T_H^2(x,Q^2,\mu _R^2)$
can be obtained from Eq.\ (\ref{eq:2.6}) using the replacement $%
x\leftrightarrow \overline{x}$ 
\begin{equation}
\label{eq:2.8}T_H^2(x,Q^2,\mu _R^2)=T_H^1(\overline{x},Q^2,\overline{\mu }%
_R^2). 
\end{equation}
The hard-scattering amplitude $T_H(x,Q^2,\mu _R^2)$ must be symmetric under
exchange $x\leftrightarrow \overline{x}$, 
\begin{equation}
\label{eq:2.9}T_H(x,Q^2,\mu _R^2)=T_H(\overline{x},Q^2,\mu _R^2). 
\end{equation}

The replacement $x\leftrightarrow \overline{x}$, by means of which the
function $T_H^2(x,Q^2,\mu _R^2)$ is found, in general, has to be applied
also to the renormalization scale $\mu _R^2$ changing it to $\overline{\mu }%
_R^2$. In the standard HSA one treats the $\mu _R^2$ and $\overline{\mu }_R^2
$ scales on the same footing by setting them equal, as a rule, to $Q^2$ .
The choice $\mu _R^2=\overline{\mu }_R^2=Q^2$ satisfies both requirements (%
\ref{eq:2.8}) and (\ref{eq:2.9}) important for the hard-scattering amplitude.
In the framework of the RC method the scales $\mu _R^2$ and $\overline{\mu }%
_R^2$ have to be chosen in accordance with Eq. (\ref{eq:2.5a}). Then the
function $T_H^1(x,Q^2,\mu _R^2=Q^2x)$ takes the following form 

\begin{equation}
\label{eq:2.11}T_H^1(x,Q^2)=\frac N{Q^2}\frac 1x\left[ 1+C_F\frac{\alpha _{%
{\rm s}}(Q^2x)}{4\pi }t(x)\right] .
\end{equation}
The second part of the hard-scattering amplitude is given by the expression 

\begin{equation}
\label{eq:2.12}T_H^2(x,Q^2)=\frac N{Q^2}\frac 1{\overline{x}}\left[ 1+C_F
\frac{\alpha _{{\rm s}}(Q^2\overline{x})}{4\pi }t(\overline{x})\right] .
\end{equation}
One can see that within the RC method the requirements (\ref{eq:2.8}) and (%
\ref{eq:2.9}) hold as well.

In the framework of both the standard HSA and RC method the $M\gamma $
transition FF can be calculated employing the formula 

$$
F_{M\gamma }(Q^2)=T_H^1(x,Q^2)\otimes \phi _M(x,Q^2)+T_H^2(x,Q^2)\otimes
\phi _M(x,Q^2)
$$

$$
=T_H^1(x,Q^2)\otimes \phi _M(x,Q^2)+T_H^1(\overline{x},Q^2)\otimes \phi
_M(x,Q^2)
$$

\begin{equation}
\label{eq:2.13}
=2T_H^1(x,Q^2)\otimes \phi _M(x,Q^2). 
\end{equation}
In the last step we take into account that the DA of the pion is a symmetric 
$\phi_M(x,Q^2)=\phi _M(\overline{x},Q^2)$ function.

The $M\equiv \pi ^0,\,\eta ,\,\eta ^{\prime }$ meson electromagnetic
transition FF's were computed within the RC method in Refs.\ \cite{ag1,ag2}.
In this work we generalize our approach by performing the computation of the 
$\pi ^0\gamma $ transition FF in the context of the RC method, but employing
instead of Eqs.\ (\ref{eq:2.11}) and (\ref{eq:2.12}) their versions symmetrized under 
$\mu _R^2\leftrightarrow \overline{\mu }_R^2$ exchange, i.e., 

\begin{equation}
\label{eq:2.14}T_H^1(x,Q^2)=\frac N{Q^2}\frac 1x\left\{ 1+C_F\frac 1{8\pi
}\left[ \alpha _{{\rm s}}(Q^2x)+\alpha _{{\rm s}}(Q^2\overline{x})\right]
t(x)\right\} ,
\end{equation}
and 

\begin{equation}
\label{eq:2.15}T_H^2(x,Q^2)=\frac N{Q^2}\frac 1{\overline{x}}\left\{
1+C_F\frac 1{8\pi }\left[ \alpha _{{\rm s}}(Q^2x)+\alpha _{{\rm s}}(Q^2
\overline{x})\right] t(\overline{x})\right\}.
\end{equation}
In the standard HSA Eqs.\ (\ref{eq:2.14}) and (\ref{eq:2.15}) coincide with
Eq.\ (\ref{eq:2.6}) and its $x\leftrightarrow \overline{x}$ partner $%
T_H^2(x,Q^2,\mu _R^2=Q^2)$, respectively. It is also not difficult to
demonstrate that Eqs.\ (\ref{eq:2.8}), (\ref{eq:2.9}), and (\ref{eq:2.13})
hold for the hard-scattering amplitude determined by the new functions $%
T_H^1(x,Q^2)$ and $T_H^2(x,Q^2)$ .

Here some comments concerning the symmetrization procedure are in order. To
clarify this important point let us note that the virtual and real photons
enter into the considering process (\ref{eq:2.2}) in an unequal manner.
Indeed, the $M\gamma $ transition FF $F_{M\gamma }(Q^2)$ depends only on $Q^2=-q_1^2$ 
($q_2^2=0$). At the same time the virtual photon-meson,

$$
\gamma ^{*}(q_1)+\gamma ^{*}(q_2)\rightarrow M(P), 
$$
transition FF $F_{M\gamma ^{*}}(Q^2,\omega )$ is a function of the photon
total virtuality Q$^2$ and asymmetry parameter $\omega $ (see the second
paper of Ref. \cite{aguila}),

$$
Q^2=Q_1^2+Q_2^2,\,\,\,\,\omega =\frac{Q_1^2}{Q^2}. 
$$
In the limits $\omega \rightarrow 0;1$ the equality%

$$
F_{M\gamma ^{*}}(Q^2,\omega =0;1)=F_{M\gamma }(Q^2) 
$$
must be valid. In order to meet this requirement and describe the real and
virtual photon-meson transitions within the RC method in a unifying way, we
adopt in this work Eqs.\ (\ref{eq:2.14}) and (\ref{eq:2.15}), because in
the limits $\omega \rightarrow 0;1$ the form factor $F_{M\gamma
^{*}}(Q^2,\omega )$ found in the context of the RC method leads to $%
F_{M\gamma }(Q^2)$, computed by means namely of Eqs.\ (\ref{eq:2.14}) and (%
\ref{eq:2.15}). The symmetrization procedure, being discussed here, was used
in Ref.\ \cite{ag5} to calculate the virtual and on-shell gluon-$\eta
^{\prime }$ meson vertex function. In the present work we concentrate on
the FF $F_{M\gamma }(Q^2)$, leaving the detailed analysis of $F_{M\gamma
^{*}}(Q^2,\omega )$ for a future publication.

\subsection{The pion distribution amplitude}

Calculation of the FF $F_{\pi \gamma }(Q^2)$ requires the knowledge of the
pion DA $\phi _\pi (x,Q^2)$, which is one of the key components in Eq.\ (\ref
{eq:2.3}). It is known \cite{cher} that the pion DA can be expanded over the
eigenfunctions of the one-loop Brodsky-Lepage equation, i.e., in terms of
the Gegenbauer polynomials $\{C_n^{3/2}(2x-1)\}$,
\begin{equation}
\label{eq:2.16}\phi _\pi (x,Q^2)=\phi _{asy}(x)\left[ 1+\sum_{n=2.4\ldots
}^\infty b_n(Q^2)C_n^{3/2}(2x-1)\right] , 
\end{equation}
where $\phi _{asy}(x)$ is the pion asymptotic DA, 
\begin{equation}
\label{eq:2.17}\phi _{asy}(x)=\sqrt{3}f_\pi x(1-x), 
\end{equation}
with $f_{\pi} =0.0923\,{\rm GeV}$ being the pion decay constant.

The evolution of the DA on the factorization scale $Q^2$ is governed by the
functions $b_n(Q^2)$,
\begin{equation}
\label{eq:2.18}b_n(Q^2)=b_n(\mu _0^2)\left[ \frac{\alpha _{{\rm s}}(Q^2)}{%
\alpha _{{\rm s}}(\mu _0^2)}\right] ^{\frac{\gamma _n}{\beta _0}}. 
\end{equation}
In Eq.\ (\ref{eq:2.18}) $\{\gamma_n\}$ are anomalous dimensions defined by the
expression
\begin{equation}
\label{eq:2.18a}
\gamma_n=C_F\left[1-\frac{2}{(n+1)(n+2)}+4\sum_{j=2}^{n+1}\frac{1}{j} \right].
\end{equation}
The constants $b_n(\mu _0^2)\equiv b_n^0$ are input parameters that form the
shape of DA's and can be extracted from experimental data or obtained from the
nonperturbative QCD computations at the normalization point $\mu _0^2$. The
QCD coupling constant $\alpha _{{\rm s}}(Q^2)$ at the two-loop approximation
are given by the expression 
\begin{equation}
\label{eq:2.19}\alpha _{{\rm s}}(Q^2)=\frac{4\pi }{\beta _0\ln (Q^2/\Lambda
^2)}\left[ 1-\frac{2\beta _1}{\beta _0^2}\frac{\ln \ln (Q^2/\Lambda ^2)}{\ln
(Q^2/\Lambda ^2)}\right] . 
\end{equation}
Here $\Lambda$ is the QCD scale parameter, $\beta _0$ and $%
\beta _1$ are the QCD beta function one- and two-loop coefficients,
respectively,
$$
\beta _0=11-\frac 23n_f,\,\,\,\,\beta _1=51-\frac{19}3n_f. 
$$

In the limit $Q^2\rightarrow \infty $ all model DA's (\ref{eq:2.16}) reduce
to the asymptotic form $\phi _{asy}(x)$. The nonasymptotic terms 
$\sim C_n^{3/2}(2x-1),\,n\geq 2$ determine the deviation of the pion DA from the
asymptotic form at moderate energy regimes and depend on the
nonperturbative mesonic binding effects.

For the pion in the literature the various phenomenological DA's were
proposed \cite{stef2,yak,cher,far,braun}. 
Thus, for example, in Ref.\ {\cite{far}}, employing the QCD\ sum rules
method, the following pion DA was predicted 
\begin{equation}
\label{eq:2.20}\phi(x,\mu _0^2)=\phi _{asy}(x)\left[
1+0.758C_2^{3/2}(2x-1)+0.3942C_4^{3/2}(2x-1)\right] ,
\end{equation}
where the normalization poin is $\mu _0=0.5\,{\rm GeV}$. 

The coefficients $b_2^0$ and $b_4^0$ were also extracted from the CLEO data on
the $\pi ^0\gamma $ transition FF in Ref.\ \cite{yak}. The authors used the
QCD light-cone sum rules  approach and included into their analysis the NLO
perturbative and twist-four corrections. They found that in the model with two
nonasymptotic terms, at the scale  $\mu _0=2.4\,{\rm GeV}$,
the pion DA has the form 
\begin{equation}
\label{eq:2.22}\phi (x,\mu _0^2)=\phi _{asy}(x)\left[
1+0.19C_2^{3/2}(2x-1)-0.14C_4^{3/2}(2x-1)\right] . 
\end{equation}
 
As is seen the pion DA's extracted from the experimental data depend on
the used methods and on their accuracy. Although one claims
that the meson DA is a process-independent quantity describing the internal
structure of the meson itself, exploration of different exclusive processes
with the same meson leads to a variety of DA's. This means that employed
methods have shortcomings or do not encompass all mechanisms important for a
given process. Such situation is pronounced in the case of the pion. The
investigation carrying out in this work intends to improve the situation
with the $\pi ^0\gamma $ transition FF by taking into account at least one
class of power corrections to the FF $F_{\pi \gamma }(Q^2)$.

To proceed it is convenient to expand the DA (\ref{eq:2.16}) over $x$ and
rewrite it in the following form: 
\begin{equation}
\label{eq:2.23}\phi _\pi (x,Q^2)=\phi _{asy}(x)\sum_{n=0}^\infty K_nx^n, 
\end{equation}
where the sum runs over all $n$. The new coefficients $K_n$ in the case of
DA's with two nonasymptotic terms are given by the expressions
$$
K_0=1+6b_2(Q^2)+15b_4(Q^2),\,\,K_1=-30[b_2(Q^2)+7b_4(Q^2)], 
$$
\begin{equation}
\label{eq:2.24}K_2=30[b_2(Q^2)+28b_4(Q^2)],\,\,K_3=-60\cdot
21b_4(Q^2),\;\;\,K_4=30\cdot 21b_4(Q^2). 
\end{equation}
Here the functions $b_2(Q^2)$ and $b_4(Q^2)$ are defined by Eq.\ (\ref
{eq:2.18}) with $\gamma_2/\beta_0$ and $\gamma_4/\beta_0$ being equal to

$$\frac{\gamma_2}{ \beta_0}=\frac{50}{81}, \;\;\; \frac{\gamma _4}{\beta
_0}=\frac{364}{405},\;\;\;n_f=3,$$
and

$$\frac{\gamma_2}{\beta_0}=\frac{2}{3},\;\;\;
\frac{\gamma_4}{\beta_0}=\frac{364}{375},\;\;\;n_f=4,$$
below and above the charm quark production threshold, respectively.

\subsection{The $\pi ^0\gamma $ transition FF within RC method}

Computation of the photon-pion transition FF $F_{\pi \gamma }(Q^2)$ implies,
naturally, integration over $x$ in accordance with Eq.\ (\ref{eq:2.13}).
Having inserted the explicit expression of the hard-scattering amplitude $%
T_H^1(x,Q^2)$ (\ref{eq:2.14}) and the pion DA\ (\ref{eq:2.23}) into Eq.\ (%
\ref{eq:2.13}) we encounter divergences, arising from the singularities
of the coupling constant $\alpha _{{\rm s}}(Q^2x)$ and $\alpha _{{\rm s}%
}(Q^2 \overline{x})$ in the limits $x\rightarrow 0;\;1$. In the standard HSA
this problem is solved by freezing the argument of the coupling constant
 and performing corresponding integrations with $%
\alpha _{{\rm s}}(Q^2)$ [or $\alpha _{{\rm s}}(Q^2/2)$]. In the RC method we
allow the QCD coupling to run and therefore have to propose some method to
cure these divergences. 

As the first step we express the running coupling 
$\alpha _{{\rm s}}(Q^2x)$\footnote {Similar consideration is valid also for
the runnig coupling $\alpha_{\rm s}(Q^2 \overline{x})$.},
in terms of $\alpha _{{\rm s}}(Q^2)$. This
aim can be achieved by applying the renormalization-group equation to $\alpha _{%
{\rm s}}(Q^2x)$ \cite{st}. As a result we find

\begin{equation}
\label{eq:2.24a}
\alpha_{\rm s}(Q^2x)\simeq \frac{\alpha_{\rm s}(Q^2)}{1+\ln x/t}
\left[1-\frac{\alpha_{\rm s}(Q^2)\beta_1}{2\pi\beta_0}\frac{\ln[1+\ln
x/t]}{1+\ln x/t} \right],
\end{equation}
where $\alpha_{\rm s}(Q^2)$ is the one-loop QCD coupling constant and
$t=4\pi / \beta_0 \alpha_{\rm s}(Q^2)=\ln \left( Q^2/\Lambda^2\right)$. Equation
(\ref{eq:2.24a}) expresses $\alpha_{\rm s}(Q^2x)$ in terms of $\alpha_{\rm
s}(Q^2)$ with an $\sim \alpha_{\rm s}^2(Q^2)$ order accuracy.

Inserting Eq.\ (\ref{eq:2.24a}) into the formula for the transition FF Eq.\
(\ref{eq:2.13}), we obtain integrals, which are still divergent, but can be
calculated using existing methods. One of them (see for details Ref.\
\cite{ag3}) allows one to obtain the form factor as a perturbative series in
$\alpha_{\rm s}(Q^2)$ with factorially growing coefficients $C_n \sim (n-1)!$,

\begin{equation}
\label{eq:2.24b}
Q^2F_{\pi \gamma}(Q^2) \sim \sum_{n=1}^{\infty}\left[ \frac{\alpha_{\rm
s}(Q^2)}{4\pi}\right]^n\beta_0^{n-1}C_n.
\end{equation}
But, it is known that a perturbative QCD series with factorially growing
coefficients is a signal for the IR renormalon nature of the divergences in
Eq.\ (\ref{eq:2.24b}). The convergence radius of such series 
is zero and its resummation should be performed by employing the Borel
integral technique. Namely, one has to determine the Borel transform
$B\left[Q^2F_{\pi\gamma} \right](u)$ of the corresponding series \cite{TZ}

\begin{equation}
\label{eq:2.24c}
B\left[Q^2F_{\pi\gamma}
\right](u)=\sum_{n=1}^{\infty}\frac{u^{n-1}}{(n-1)!}C_n,
\end{equation}
and in order to define the sum Eq.\ (\ref{eq:2.24b}), or to find the
resummed expression for the form factor, one has to invert
$B\left[Q^2F_{\pi\gamma} \right](u)$ to get

\begin{equation}
\label{eq:2.24d}
\left[Q^2F_{\pi\gamma}(Q^2) \right]^{res} \sim
{\rm P.V.} \int_0^{\infty}du \exp \left[-\frac{4\pi u}{\beta_0\alpha_{\rm s}(Q^2)}\right]
B\left[ Q^2F_{\pi\gamma}\right](u).
\end{equation}
Because the coefficients of the series Eq.\ (\ref{eq:2.24b}) behave like
$C_n \sim (n-1)!$, the Borel transform (\ref{eq:2.24c}) contains poles
located at the positive $u$ axis of the Borel plane, which  are exactly the IR
renormalon poles. Therefore the inverse Borel transformation
(\ref{eq:2.24d}) can be computed only after regularization of these pole
singularities. One of the methods of such regularization, adopted also in the
present work, is the principal
value prescription. In other words, the IR renormalon divergences in Eq.\
(\ref{eq:2.24d}) have to be removed by computing the integral in the sense
of the  Cauchy principal value. Only after this regularization the inverse
Borel transformation defines the resummed FF.

Fortunately, these intermediate operations can be omitted with the help of
the following operations. Namely, let us introduce the inverse Laplace
transformations of the functions in Eq.\ (\ref{eq:2.24a}), i.e.,

\begin{equation}
\label{eq:2.24f}
\frac{1}{(t+z)^\nu}=\frac{1}{\Gamma(\nu)}\int_0^{\infty}du
\exp[-u(t+z)]u^{\nu-1}, \;\; Re\nu>0,
\end{equation}
and

\begin{equation}
\label{eq:2.24m}
\frac{\ln [t+z]}{(t+z)^2}=\int_0^{\infty}du\exp[-u(t+z)](1-\gamma_E- \ln u)u,
\end{equation}
where $\Gamma(z)$ is the Gamma function, $\gamma_E \simeq 0.577216$ is the
Euler constant, and $z=\ln x$ [or $z= \ln \overline{x}$ in the case of
$\alpha_{\rm s}(Q^2\overline{x})$]. Then, using Eqs.\
(\ref{eq:2.24f}) and (\ref{eq:2.24m}) for the QCD coupling $\alpha_{\rm
s}(Q^2x)$ we find \cite{ag2,ag5}

\begin{equation}
\label{eq:2.25}\alpha _{{\rm s}}(Q^2x)=\frac{4\pi }{\beta _0}%
\int_0^\infty due^{-ut}R(u,t)x^{-u}, 
\end{equation}
where the function $R(u,t)$ is defined as

\begin{equation}
\label{eq:2.26}R(u,t)=1-\frac{2\beta _1}{\beta _0^2}u(1-\gamma _E-\ln t-\ln
u).
\end{equation}

Having used Eq.\ (\ref{eq:2.25}) and performed integration over $x$, for the
scaled $\pi ^0\gamma $ transition FF we get 
$$
Q^2F_{\pi \gamma }(Q^2)=\sqrt{3}f_\pi N\left\{ \sum_{n=0}^\infty \frac{K_n}{%
n+1}+\frac 4{3\beta _0}\int_0^\infty due^{-ut}R(u,t)\right. 
$$
\begin{equation}
\label{eq:2.27}\left. \times \sum_{n=0}^\infty K_n\left[ A_n(u)+\tilde
A_n(u)\right] \right\} . 
\end{equation}
Here the term $\sim A_n(u)$ appears in the result of the integration of the
second term in Eq.\ (\ref{eq:2.14}), whereas the term $\sim \tilde A_n(u)$ 
owing to the third term in Eq.\ (\ref{eq:2.14}). The functions $A_n(u)$ and 
$\tilde A_n(u)$ have the following forms: 
$$
A_n(u)=\frac{d^2}{d\beta ^2}B(2,\beta )|_{\beta =n+1-u}-\frac d{d\beta
}B(1,\beta )|_{\beta =n+2-u}-9B(2,n+1-u) 
$$
\begin{equation}
\label{eq:2.28}
=\frac 2{(n+1-u)^3}-\frac 2{(n+2-u)^3}+\frac
1{(n+2-u)^2}-\frac 9{(n+1-u)(n+2-u)}, 
\end{equation}
and
$$
\tilde A_n(u)=\frac{\partial ^2}{\partial \beta ^2}B(2-u,\beta )|_{\beta
=n+1}-\frac \partial {\partial \beta }B(1-u,\beta )|_{\beta
=n+2}-9B(2-u,n+1) 
$$
$$
=B(n+1,2-u)\left[ \left( \psi (n+1)-\psi (n+3-u)\right) ^2+\psi ^{\prime
}(n+1)-\psi ^{\prime }(n+3-u)\right] 
$$
\begin{equation}
\label{eq:2.29}
-B(n+2,1-u)\left[ \psi (n+2)-\psi (n+3-u)\right] -9B(2-u,n+1), 
\end{equation}
where $B(x,y)$ is the beta function $B(x,y)=\Gamma (x)\Gamma (y)/\Gamma
(x+y) $ and $\psi (z)=d[\ln \Gamma (z)]/dz$.

The functions $A_n(u)$ and $\tilde A_n(u)$ contain the poles on the positive
real axis of the plane $u$. Indeed, the function $A_n(u)$ has the finite
number of triple, double, and single poles located at the points $u_0=n+2$
and triple and single ones at  $u_0=n+1$. In order to reveal the pole
structure of the function $\tilde A_n(u)$, it is convenient to use the
following formulas \cite{prud} 
\begin{equation}
\label{eq:2.30}\psi (z)=-\gamma _E+(z-1)\sum_{k=0}^\infty \frac
1{(k+1)(k+z)},\;\psi ^{\prime }(z)=\sum_{k=0}^\infty \frac 1{(k+z)^2}.
\end{equation}
Here we write down, as an example, the function $\tilde A_0(u)$, which
after some manipulations takes the form 
$$
\widetilde{A}_0(u)=(2-u)\left[ \sum_{k=0}^\infty \frac
1{(k+1)(k+3-u)}\right] ^2-\frac 1{2-u}\sum_{k=0}^\infty \frac 1{(k+3-u)^2} 
$$
\begin{equation}
\label{eq:2.31}+\frac 1{1-u}\sum_{k=0}^\infty \frac 1{(k+1)(k+3-u)}-\frac
1{1-u}+\frac{\psi ^{\prime }(1)-8}{2-u}.
\end{equation}
Now it is clear that $\tilde A_0(u)$ contains the infinite number of the
double poles located at $u_0=\,k+3$ and the single ones at $u_0=1,\,2,\,k+3$.
 The similar analysis can be fulfilled for $\tilde A_n(u),\,n>0$ as well.
Hence by employing Eq.\ (\ref{eq:2.25}) we have transformed the end-point $%
x\rightarrow 0;\,1$ divergences in Eq.\ (\ref{eq:2.13}) into the IR
renormalon pole
divergences in Eq. (\ref{eq:2.27}). The integral in Eq.\ (\ref{eq:2.27}) is the
inverse Borel transformation (\ref{eq:2.24d}), where the Borel transform $B[Q^2F_{\pi \gamma
}](u)$ of the NLO part of the scaled FF is defined (up to constant factor)
as 
\begin{equation}
\label{eq:2.32}B[Q^2F_{\pi \gamma }](u)\sim R(u,t)\sum_{n=0}^\infty
K_n\left[ A_n(u)+\tilde A_n(u)\right] .
\end{equation}
The IR renormalon divergences in Eq.\ (\ref{eq:2.27}) must be removed
by means of the the principal value prescription. 
The inverse Borel transformation after such regularization, as we have just
pointed out above,  becomes the resummed form factor $[Q^2F_{\pi \gamma }(Q^2)]^{res}$.
Therefore all integrals over $u$ hereafter have to be understood in the
sense of the Cauchy principal value.

The expression $[Q^2F_{\pi \gamma }(Q^2)]^{res}$ contains the power-suppressed 
corrections $\sim 1/Q^{2n},\,\,n=1,2, \ldots$ to the scaled FF, implicitly
existing in the QCD\ factorization formula (\ref{eq:2.3}). The detailed discussion 
of relevant problems can be found in
Refs.\ \cite{ag2,ag5}. Here, for completeness, we outline 
the important points of this analysis. To make the discussion of this question
as transparent as possible, let us for a moment neglect the nonleading term $\sim
\alpha_{\rm s}^2(Q^2)$ in Eq.\ (\ref{eq:2.24a}) and consequently make the replacement
$R(u,t) \to 1$ in Eq.\ (\ref{eq:2.25}). Then the integrals in the scaled and
resummed FF with multiple IR renormalon poles at $u_0=n$ can be easily
expressed in terms of the integrals with a single IR renormalon pole at the
same point [see Eqs.\ (\ref{eq:3.6}) and (\ref{eq:3.15})], 
so that our formula (\ref{eq:2.27}) will consist of some 
linear combinations of the integrals,

\begin{equation}
\label{eq:2.32a}
\frac{4\pi}{\beta_0}\int_0^{\infty}\frac{e^{-ut}du}{n-u}=\frac{1}{n}f_{2n}(Q),
\end{equation}
where $f_{2n}(Q)$ are the moment integrals,

\begin{equation}
\label{eq:2.32b}
f_p(Q)=\frac{p}{Q^p}\int_0^{Q}dkk^{p-1}\alpha_{\rm s}(k^2).
\end{equation}
The integrals $f_p(Q)$ were calculated in Ref.\ \cite{W98} using the IR
matching scheme:

\begin{equation}
\label{eq:2.32c}
f_p(Q)=\left( \frac{\mu_I}{Q} \right)^pf_p(\mu_I)+\alpha_{\rm
s}(Q^2) \sum_{n=0}^N \left[ \frac{\beta_0}{2 \pi p} \alpha_{\rm
s}(Q^2) \right ]^n[n!-\Gamma(n+1,\;p\ln (Q/\mu_I))],
\end{equation}
where $\mu_I$ is the infrared matching scale and $\Gamma(n+1,\;z)$ is the
incomplete Gamma function. In Eq.\ (\ref{eq:2.32c}) $\{ f_p(\mu_I) \}$ are
phenomenological parameters, which represent the weighted average of
$\alpha_{\rm s}(k^2)$ over the IR region $0<k<\mu_I$ and act at the same
time as infrared regulators of the right-hand side (RHS) of Eq.\ (\ref{eq:2.32a}). 
The first term on
the right-hand side of Eq.\ (\ref{eq:2.32c}) is the power-suppressed contribution to
$f_p(Q)$ and models the "soft" part of the moment integral. It cannot be
calculated within the perturbative QCD, whereas the second term on the RHS
of Eq.\ (\ref{eq:2.32c}) is the perturbatively calculable part of the
function $f_p(Q)$, representing its hard perturbative "tail." In other
words, the IR matching scheme allows one to estimate power corrections to
the moment integrals by explicitly dissecting them out from the full
expression, and introducing new nonperturbative parameters $f_p(\mu_I)$. The
same moment integrals $f_p(Q)$ computed in the framework of the RC method
[LHS of Eq.\ (\ref{eq:2.32a})], contain information on both their soft and
the perturbative components. Therefore we can state that the scaled and
resummed FF (\ref{eq:2.27}) contain power corrections $\sim 1/Q^{2n}$. In
phenomenological applications both the IR matching scheme and the RC method
can be employed. But the RC method has an advantage over the IR matching
scheme, because it allows one to compute the functions $f_p(Q)$ without
introducing the new nonperturbative parameters $\mu_I$ and $f_p(\mu_I)$.
Moreover, using this method, the parameters $f_p(\mu_I)$ themselves can be
calculated in good agreement with model calculations and available
experimental data \cite{ag5,ag6}.

But the principal
value prescription itself generates in the each integral over $u$
higher-twist  ambiguities, 
$$
\sim \sum_qN_q\frac{\Phi _q(Q^2)}{Q^{2q}}, 
$$
where $\Phi _q(Q^2)$ is a calculable function fixed by the residue of the
integral at the pole $u_0=q$ and $N_q$ is some numerical constant. The
ambiguities taken into account in Eq. (\ref{eq:2.27}) modify the Borel
resummed $\pi ^0\gamma $ transition FF, yielding 
\begin{equation}
\label{eq:2.33}[Q^2F_{\pi \gamma }(Q^2)]^{res}\rightarrow [Q^2F_{\pi \gamma
}(Q^2)]^{res}+[Q^2F_{\pi \gamma }(Q^2)]^{HT}.
\end{equation}
The HT term depends on the known functions $\{\Phi _q(Q^2)\}$ and
coefficients $\{K_q\}$ and on the unknown numerical constants $\{N_q\}$. In
accordance with the ''ultraviolet dominance assumption'' this HT ambiguity
allows one to estimate higher-twist corrections to the scaled form factor $%
Q^2F_{\pi \gamma }(Q^2)$ coming from sources another than end-point
integration. By fitting the constants $\{N_q\}$ to experimental data one can
deduce some information concerning the magnitude of such corrections.

\section{Asymptotic limit of the resummed $\pi ^0\gamma $ transition FF}

\setcounter{equation}0

As we have emphasized above, the resummed $\pi ^0\gamma $ transition FF
contains the power corrections appearing due to the end-point
integration. These corrections in the region of a moderate momentum transfer 
$Q^2$ are essential for explaining the experimental data \cite
{ag1,ag2,ag4,ag7}. But it is also evident that in the asymptotic limit $%
Q^2\rightarrow \infty $ , where all higher-twist corrections should vanish,
the standard HSA with frozen $\alpha _{{\rm s}}(Q^2)$ and the pion
asymptotic DA $\phi _{asy}(x)$ leads to the correct expression for the $\pi
^0\gamma $ transition FF. Consequently,
in the limit $Q^2\rightarrow \infty $ from the resummed FF we have to regain
the asymptotic one.

In the limit $Q^2\rightarrow \infty $ the pion DA goes to its asymptotic
form, i.e., 
\begin{equation}
\label{eq:3.1}\phi _\pi (x,Q^2)\stackrel{Q^2\rightarrow \infty }{%
\longrightarrow }\phi _{asy}(x), 
\end{equation}
which in the terms of the coefficients $K_n$ means 
\begin{equation}
\label{eq:3.2}K_0\rightarrow 1,\,\,\,\,K_n\rightarrow 0,\,\,n>0. 
\end{equation}
We take also into account that in this limit the subleading term in the
expansion of $\alpha _{{\rm s}}(Q^2x)$ through $\alpha _{{\rm s}%
}(Q^2) $ has to be neglected. In other words, in the limit $Q^2\rightarrow
\infty $ we have to fulfill the replacement 
\begin{equation}
\label{eq:3.3}\int_0^\infty due^{-ut}R(u,t)\stackrel{Q^2\rightarrow \infty }{%
\longrightarrow }\int_0^\infty due^{-ut}. 
\end{equation}
After these operations the resummed FF takes the following form: 
\begin{equation}
\label{eq:3.4}[Q^2F_{\pi \gamma }(Q^2)]^{res}\stackrel{Q^2\rightarrow \infty 
}{\longrightarrow }\sqrt{3}f_\pi N\left\{ 1+\frac 4{3\beta _0}\int_0^\infty
due^{-ut}\left[ A_0(u)+\widetilde{A}_0(u)\right] \right\} . 
\end{equation}
But Eq. (\ref{eq:3.4}) is not the final expression, because in the integral
above $t=\ln (Q^2/\Lambda ^2)$ and its $Q^2\rightarrow \infty $ limit still
has to be computed.

To this end, we start from the simple case and consider the integral 
\begin{equation}
\label{eq:3.5}I_1=\int_0^\infty due^{-ut}A_0(u)=\int_0^\infty
due^{-ut}\left[ \frac 2{(1-u)^3}-\frac 2{(2-u)^3}+\frac 1{(2-u)^2}-\frac
9{1-u}+\frac 9{2-u}\right].
\end{equation}
The integrals in Eq. (\ref{eq:3.5}) with the triple and double poles can be
reduced to ones with single poles:
$$
\int_0^\infty \frac{e^{-ut}du}{(n-u)^3}=-\frac 1{2n^2}-\frac{\ln \lambda }{2n%
}+\frac{\ln {}^2\lambda }2\frac{li(\lambda ^n)}{\lambda ^n}, 
$$
\begin{equation}
\label{eq:3.6}\int_0^\infty \frac{e^{-ut}du}{(n-u)^2}=-\frac 1n+\ln
{}\lambda \frac{li(\lambda ^n)}{\lambda ^n},\,\,\,\,\,\int_0^\infty \frac{%
e^{-ut}du}{n-u}=\frac{li(\lambda ^n)}{\lambda ^n},
\end{equation}
where the first two equalities are obtained performing integrations by
parts. Here the logarithmic integral $li(\zeta)$ is defined as 
\begin{equation}
\label{eq:3.7}li(\zeta)={\rm P.V.}\int_0^{\zeta}\frac{dt}{\ln t},
\end{equation}
and $\lambda =Q^2/\Lambda ^2$. Now using the expansion of $li(\zeta^n)/\zeta^n$ in
inverse powers of $\ln \zeta$ \cite{ag5}, 
\begin{equation}
\label{eq:3.8}\frac{li(\zeta^n)}{\zeta^n}\simeq \frac 1{n\ln \zeta}\sum_{m=0}^M\frac{m!}{%
(n\ln \zeta)^m},\,\,M\gg 1,
\end{equation}
and keeping in the expressions 
$$
\ln {}^2\lambda \frac{li(\lambda ^n)}{\lambda ^n},\,\,\,\,\,\ln {}\lambda 
\frac{li(\lambda ^n)}{\lambda ^n}, 
$$
terms up to $O(1/\ln \lambda )$ order, we get 
\begin{equation}
\label{eq:3.9}I_1\stackrel{Q^2\rightarrow \infty }{\longrightarrow }-\frac
52\frac 1{\ln \lambda }.
\end{equation}

The situation with the second integral, 
\begin{equation}
\label{eq:3.10}I_2=\int_0^\infty due^{-ut}\widetilde{A}_0(u),
\end{equation}
is more subtle. In this case, instead of using the explicit form of 
$\widetilde{A}_0(u)$, we consider the integral 
\begin{equation}
\label{eq:3.11}I_2=\int_0^\infty due^{-ut}\int_0^1dx \overline{x}%
^{1-u}t(x)=\int_0^\infty due^{-ut}\int_0^1dx x^{1-u}t(\overline{x}),
\end{equation}
from which Eq.\ (\ref{eq:3.10}) has been derived. We are going to explain our
technique, analyzing one of the components of the function $t(\overline{x})$.
Namely, let us calculate the $Q^2\rightarrow \infty $ limit of the
integral 
\begin{equation}
\label{eq:3.12}I_2^1=\int_0^\infty due^{-ut}\int_0^1dxx^{1-u}\ln {}^2(1-x).
\end{equation}
Having expanded $\ln {}^2(1-x)$ in powers of $x$,
\begin{equation}
\label{eq:3.13}\ln {}^2(1-x)=2\sum_{k=1}^\infty \frac 1{k+1}[\psi (k+1)-\psi
(1)]x^{k+1},
\end{equation}
we obtain 
\begin{equation}
\label{eq:3.14}\int_0^1dxx^{1-u}\ln {}^2(1-x)=2\sum_{k=1}^\infty \frac
1{k+1}[\psi (k+1)-\psi (1)]\frac 1{k+3-u}.
\end{equation}
Substituting Eq. (\ref{eq:3.14}) into the integral $I_2^1$,
\begin{equation}
\label{eq:3.15}2\sum_{k=1}^\infty \frac 1{k+1}[\psi (k+1)-\psi
(1)]\int_0^\infty \frac{due^{-ut}}{k+3-u}=2\sum_{k=1}^\infty \frac
1{k+1}[\psi (k+1)-\psi (1)]\frac{li(\lambda ^{k+3})}{\lambda ^{k+3}},
\end{equation}
and using the leading order term in expansion (\ref{eq:3.8}), in the limit $%
Q^2\rightarrow \infty $ we get 
\begin{equation}
\label{eq:3.16}I_2^1\stackrel{Q^2\rightarrow \infty }{\longrightarrow }%
\,\,2\sum_{k=1}^\infty \frac 1{k+1}[\psi (k+1)-\psi (1)]\frac 1{k+3}\frac
1{\ln \lambda }.
\end{equation}
Now, having repeated the described above operations in the reverse order, it
is easy to see that 
\begin{equation}
\label{eq:3.17}\,2\sum_{k=1}^\infty \frac 1{k+1}[\psi (k+1)-\psi (1)]\frac
1{k+3}\frac 1{\ln \lambda }=\frac 1{\ln \lambda }\int_0^1dxx\ln {}^2(1-x).
\end{equation}
In other words, the asymptotic limit $Q^2\rightarrow \infty $ transforms the
integral $I_2^1$ in accordance with the rule 
\begin{equation}
\label{eq:3.18}I_2^1\stackrel{Q^2\rightarrow \infty }{\longrightarrow }%
\,\,\,\frac 1{\ln \lambda }\int_0^1dxx\ln {}^2(1-x).
\end{equation}
The same conclusion is valid also for the other terms from Eq. (\ref{eq:2.7}%
). Summing up, we derive the limit of the integral $I_2$,
$$
I_2\stackrel{Q^2\rightarrow \infty }{\longrightarrow }\,\,\,\frac 1{\ln
\lambda }\int_0^1dx xt(1-x)=\frac 1{\ln \lambda }B(1,2)\left\{ \left[ (\psi
(1)-\psi (3))^2+\psi ^{\prime }(1)-\psi ^{\prime }(3)\right] \right.  
$$
\begin{equation}
\label{eq:3.19}\left. - \left[ \psi (2)-\psi (3)\right] -9\right\} .
\end{equation}
The expression (\ref{eq:3.19}) without the factor $1/\ln \lambda$ is nothing more 
than $\widetilde{A}_0(u)$ at $u=0$. The following operations are trivial and lead to 
\begin{equation}
\label{eq:3.20}I_2\stackrel{Q^2\rightarrow \infty }{\longrightarrow }%
\,-\frac 52\frac 1{\ln \lambda }.
\end{equation}
As is seen, both the functions $I_1$ and $I_2$ have the same limits.
Consequently, for the resummed FF we find 
\begin{equation}
\label{eq:3.21}[Q^2F_{\pi \gamma }(Q^2)]^{res}\stackrel{Q^2\rightarrow
\infty }{\longrightarrow }\,2f_\pi \left[ 1-\frac 5{3\pi }\alpha _{{\rm s}%
}(Q^2)\right] ,
\end{equation}
in deriving of which the value of the constant $N$, 
\begin{equation}
\label{eq:3.22}N=\sqrt{12}(e_u^2-e_d^2)
\end{equation}
for the pion has been utilized. Equation (\ref{eq:3.21}) can be readily
obtained within the standard HSA employing the pion asymptotic DA. Our
analysis proves that in the asymptotic limit the Borel resummed FF leads to
the correct expression (\ref{eq:3.21}), which we consider as one of
justifications of the symmetrization procedure. It is also worth remarking
that the ''old'' version of the hard-scattering amplitude $T_H(x,Q^2)$ [see
Eqs. (\ref{eq:2.11}) and (\ref{eq:2.12})] gives the correct asymptotic FF as
well, that is evident from Eq. (\ref{eq:3.9}). Hence in the asymptotic
limit both the ordinary and symmetrized RC\ methods describe correctly the 
$\pi ^0\gamma $ transition FF, the difference between them being sizeable at the
moderate values of the momentum transfer, $Q^2\sim $a few ${\rm GeV}^2$.

\section{Extracting the pion DA from the CLEO data}

\setcounter{equation}0

In this section we present the pion phenomenological DA's extracted from the
CLEO\ data within the SRC method. In our calculations below we
shall use the following values of the parameters $\Lambda$ and $\mu_0$

\begin{equation}
\label{eq:4.1}
\Lambda_4=0.25\ {\rm GeV}, \;\;\; \mu_0^2=1\ {\rm GeV}^2.
\end{equation}

As is known (see for review Ref.\ \cite{ben}), the IR renormalon calculus
can be applied for the estimation of power corrections to some physical
quantity in the region of the high momentum transfers $Q^2 \gg \Lambda^2$.
Our choice for the parameters (\ref{eq:4.1}) leads to the requirement $16\ Q^2
\gg 1$. Because the recent CLEO data \cite{cleo} on the $\pi^0\gamma$
transition FF lie in the domain $1.64\ {\rm GeV}^2 \leq Q^2 < 10\ {\rm
GeV}^2$, we include them into our numerical analysis to deduce the pion
model DA's. Namely at these moderate momentum transfers the power
corrections play the important role, modifying both quatitatively and qualitatively
predictions for $Q^2F_{\pi\gamma}(Q^2)$ obtained within the standard HSA.

The Borel resummed $\pi ^0\gamma $ transition FF implies summations over $n$
and $k$, the latter arising from $\sim \widetilde{A}_n(u)$ terms. The
summation over $n$ does not create problems, because in our studies we use
the pion asymptotic and model DA's with one and two nonasymptotic terms.
Therefore the maximal value of $n$ in the sum is $N_{\max }=0$ and $2,\;4$,
respectively. It is worth noting that Eq. (\ref{eq:2.27}) is a general
expression valid for the pion DA's with an arbitrary number of nonasymptotic
terms. The next terms $\sim C_n(2x-1),\,n>4$ can be easily included into our
scheme by modifying only expressions of the coefficients $K_n$ and $N_{\max
} $. Contrary to the case with $n$, at fixed $n$ summation over $k$ runs
from $k=0$ to $k=\infty $ and has to be truncated at some $k_{\max }$. In
other words, the results of numerical computations depend on $k_{\max }$. In
order to check their sensitivity to a chosen value of $k_{\max }$, we have
performed calculation of the FF with $k_{\max }=50$ and $k_{\max }=100$. We
have found that for the pion asymptotic DA the ratio 
\begin{equation}
\label{eq:4.2}R(Q^2)=\frac{[Q^2F_{\pi \gamma }(Q^2)]^{res}(k_{\max }=100)}{%
[Q^2F_{\pi \gamma }(Q^2)]^{res}(k_{\max }=50)} 
\end{equation}
at $Q^2=1\,{\rm GeV}^2$ is equal to $R(1\,{\rm GeV}^2)=1.0027$ and to $R(10\,%
{\rm GeV}^2)=1.0011$ at $Q^2=10\,{\rm GeV}^2$. We have obtained a similar
picture employing the pion various model DA's. Since the correction to $%
Q^2F_{\pi \gamma }(Q^2)$ originating from the next $k=51-100$ terms does not
exceed $3\cdot 10^{-3}$ of those from the first $k=0-50$ ones, in numerical
computations we set $k_{\max }=50$. Such output is understandable, because
in the resummed FF dominate contributions from the nearest to $u=0$ IR
renormalon poles.

\begin{figure}
\epsfxsize=10 cm
\epsfysize=8 cm
\centerline{\epsffile{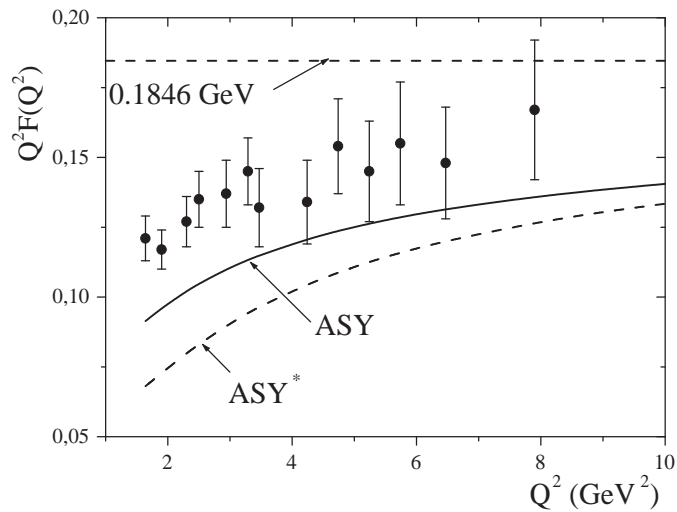}}
\caption{\label{fig:fig1} The scaled and resummed $\pi ^0\gamma $ transition
form factor $Q^2F_{\pi
\gamma }(Q^2)$ as a function of $Q^2$. The curves ASY and ASY$^{*}$
are computed using the pion asymptotic DA (\ref{eq:2.17}). The curve ASY$^{*}$
corresponds to the FF obtained within the ordinary RC approach, whereas for
calculation of the curve ASY the SRC method is employed. The upper
dashed line shows the
model-independent $Q^2\rightarrow \infty $ limit for the FF. The data are
borrowed from Ref. \cite{cleo}.}
\end{figure}

We start our analysis of the $\pi ^0\gamma $ transition FF from the
pion asymptotic DA in order to reveal the impact of the symmetrization
procedure on the predictions, as well as to find out how large is
the deviation of these predictions from the data points. Our results are shown
in Fig.\ \ref{fig:fig1}. As is seen, the scaled and resummed FF $Q^2F_{\pi\gamma}(Q^2)$
computed using the SRC method in the region of the momentum
transfers $1.64\ {\rm GeV}^2 \leq Q^2 \leq 5\ {\rm GeV}^2$ are considerably 
larger than the one obtained by means of the ordinary
RC approach. As a result, the deviation of the curve ASY from the data
points are smaller than that of ASY$^{*}$. Nevertheless, such deviation exists and some
admixture of nonasymptotic terms in the pion DA is needed to explain the
data. 

\begin{figure}
\epsfxsize=10 cm
\epsfysize=8 cm
\centerline{\epsffile{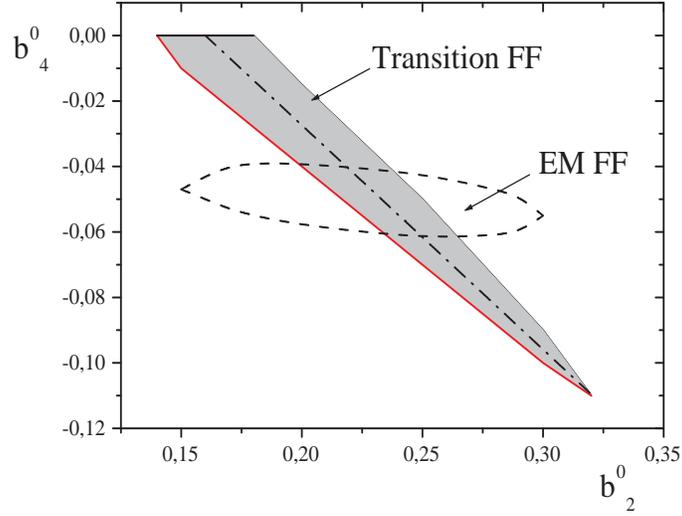}}
\caption{\label{fig:fig2}The $1\sigma$ areas in the $b_4^0-b_2^0$ plane of
the input
parameters. The shaded area is found from analysis of the CLEO data on the
$\pi^0\gamma$ transition FF. The region bounded by the dashed lines is
extracted from the data on the pion electromagnetic (EM) FF. The dot-dashed line
is the diagonal determined by Eq.\ (\ref{eq:4.4}).}
\end{figure}

In Fig.\ \ref{fig:fig2} we depict (the shaded area) the $1\sigma$ area for
values of the input
parameters $b_2^0,\;b_4^0$ in the $b_4^0-b_2^0$ plane. This means that the
$\pi^0\gamma$ transition FF computed in the context of the SRC
method employing the pion model DA's with Gegenbauer coefficients
belonging to the shaded region describes the CLEO data with a $1\sigma$
accuracy.

\begin{figure}
\epsfxsize=10 cm
\epsfysize=8 cm
\centerline{\epsffile{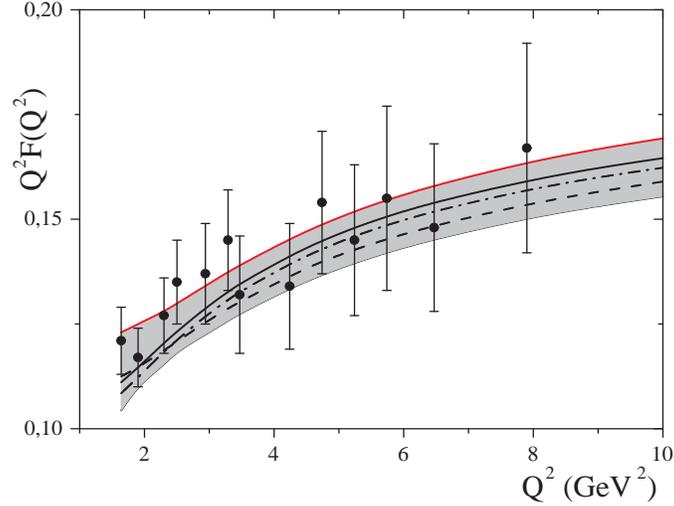}}
\caption{\label{fig:fig3}  The scaled and resummed $\pi ^0\gamma $
transition FF vs $Q^2$.
The shaded area demonstrates $1\sigma$ region for the FF. Correspondence
between the curves and the input parameters is; for the solid line
$b_2^0=0.25,\; b_4^0=-0.05$; for the dashed line $b_2^0=0.16,\;b_4^0=0$; and
for the dot-dashed line $b_2^0=0.23, \; b_4^0=-0.05$.}
\end{figure}

In Fig.\ \ref{fig:fig3} we plot the $1\sigma$ area for the $\pi^0\gamma$ transition FF
itself. The central curves with $b_2^0=0.16,\;b_4^0 =0$ (the DA's with one
nonasymptotic term) and $b_2^0=0.25,\;b_4^0 =-0.05$ (the DA's with two
nonasymptotic terms) are also shown. One sees that the shapes of the curves
with $b_4^0 =0$ and $b_4^0 < 0$ differ from each other. Indeed, the curves with
$b_4^0 <0$ are sharper relative to ones with $b_4^0 =0$. Therefore the
boundaries of the $1\sigma$ area are determined by superposition of curves of these two
types.
Our analysis in the case of the DA's with one nonasymptotic term leads to
the following estimation

\begin{equation}
\label{eq:4.3}
b_2^0=0.16 \pm 0.02, \;\;\;b_4^0=0.
\end{equation}
In the case of DA's with two nonasymptotic terms allowed valus of
$b_2^0,\;b_4^0$ cover the shaded area in Fig.\ \ref{fig:fig2}. Below we write down
sample values of the parameters,

$$
b_2^0=0.2,\;\;\;b_4^0 \in [ -0.02,\ -0.04],
$$
$$
b_2^0=0.25,\;\;\;b_4^0 \in [-0.05,\ -0.07],
$$
and
$$
b_2^0=0.3,\;\;\; b_4^0 \in [-0.09,\ -0.1].
$$
The "diagonal" of the $1\sigma$ area is determined by the expression

\begin{equation}
\label{eq:4.4}
b_2^0+1.46b_4^0=0.16,\;\;b_4^0 \in [0, -0.11]
\end{equation}
and is shown in Fig.\ \ref{fig:fig2} by the dot-dashed line.

\begin{figure}
\epsfxsize=10 cm
\epsfysize=8 cm
\centerline{\epsffile{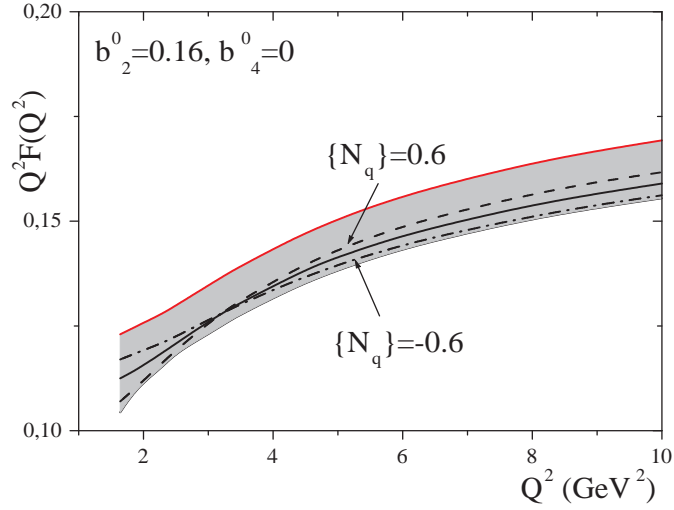}}
\caption{\label{fig:fig4} The scaled and resummed form factor
$Q^2F_{\pi\gamma}(Q^2)$
with and without HT ambiguities. The solid line describes FF without HT
ambiguities. The broken lines are found employing Eq.\ (\ref{eq:2.33}) and
numerical constants $\{N_q\}$ shown in the figure.}
\end{figure}

As we have noted above, the principal value prescription generates HT
ambiguities that in conjunction with the ''ultraviolet dominance
assumption'' can be used to estimate HT corrections to the form factor
originating from another source (for example, from the pion HT DA's). We
have performed relevant computations, as a sample, for the pion DA with one
nonasymptotic term $b_2^0(1\,\,\,{\rm GeV}^2)=0.16$ and $\{N_q\}=\pm
0.6,\;q=1,2,\ldots, 50$
(Fig.\ \ref{fig:fig4}). We find that the values $\{N_q\}=\pm 0.6$
determine the upper and lower bounds for the constants $\{N_q \}$ in order
that FF remain within the $1\sigma$ region. The $\pi^0\gamma$ transition FF
with HT ambiguities corresponding to $\{N_q\}=-0.6$ ($\{N_q\}=0.6$) at $Q^2
< 3.5\ {\rm GeV}^2$ is larger (smaller) than the FF without such corrections
and are smaller (larger) for $Q^2 > 3.5\ {\rm GeV}^2$. The HT 
ambiguities, obeying the "$1\sigma$ constraint" do not exceed the level $\sim
\pm 5 \%$ of the transition FF at the momentum transfers 
$Q^2=1.64-2 \; {\rm GeV}^2$ and reach only $\sim \mp 1.8\%$ at $Q^2=9-10 \; {\rm GeV}^2$.

\section{The electromagnetic form factor $F_\pi (Q^2)$}

\setcounter{equation}0

In this section we compute the pion electromagnetic FF $F_\pi (Q^2)$ in the
framework of the RC method in order to extract further constraints on the
parameters $b_2^0,\;b_4^0$. 

The FF $F_\pi (Q^2)$ is
the important quantity characterizing the pion, which was thoroughly
investigated in the context of PQCD \cite{gupta,li,ag6}. It was also studied
in various experiments \cite{bebek,volmer}. Within the RC method FF's 
$F_M(Q^2)$ of the light mesons $M=\pi ,\,K,\,\rho _L$ were considered in 
Refs.\ \cite{ag3,ag4,ag7,ag6}. Therefore below we outline only main
stages of the RC analysis of the FF $F_M(Q^2)$.

In the standard HSA\ the FF $F_\pi (Q^2)$ is given by the factorization
formula 
\begin{equation}
\label{eq:5.1}F_\pi (Q^2)=\int_0^1dx\int_0^1dy\phi _\pi ^{*}(y,Q^2)T_H\left(
x,y,Q^2\right) \phi _\pi (x,Q^2). 
\end{equation}
Here $T_H(x,y,Q^2)$ is the hard-scattering amplitude of the subprocess $q 
\overline{q}^{\prime }+\gamma ^{*}\rightarrow q\overline{q}^{\prime }$, $%
Q^2=-q^2$ is the momentum transfer, $q$ being the four-momentum of the
virtual photon. In Eq.\ (\ref{eq:5.1}) the factorization scale from the very
beginning is chosen equal to $\mu _F^2=Q^2$.

At the leading order of PQCD the amplitude $T_H(x,y,Q^2)$ has the form 
\begin{equation}
\label{eq:5.2}T_H(x,y,Q^2)=\frac{16\pi C_F}{Q^2}\left[ \frac 23\frac{\alpha
( \overline{\mu }_R^2)}{\overline{x}\overline{y}}+\frac 13\frac{\alpha (\mu
_R^2)}{xy}\right] . 
\end{equation}

In accordance with the ideology of the RC method the argument of the QCD\
coupling in Eq.\ (\ref{eq:5.2}) has to be chosen as 
\begin{equation}
\label{eq:5.3}\mu _R^2=xyQ^2,\;\;\overline{\mu }_R^2=\overline{x}\overline{y}%
Q^2. 
\end{equation}
Such choice allows one to get rid of terms $\sim \ln (\overline{x}\overline{y
}Q^2/\overline{\mu}_R^2),\,\ln (xyQ^2/\mu_R^2)$ appearing in the amplitude $T_H(x,y,Q^2)$ at the
next-to-leading order of PQCD and minimizes the higher-order corrections to $%
F_\pi (Q^2)$. We can also adopt the scheme 
\begin{equation}
\label{eq:5.4}\mu _R^2=xQ^2,\;\;\overline{\mu }_R^2=\overline{x}Q^2, 
\end{equation}
obtained from Eq. (\ref{eq:5.3}) by freezing $y$. In the framework of the
standard HSA one freezes both of $x,\,$ $y$ and compute the form factor with 
$\mu _R^2=\overline{\mu }_R^2=Q^2$ (or $Q^2/4$ ) \cite{gupta}.

In the above we have chosen $x$ as the running variable. Alternatively, we can
fix $x$ and choose $y$ as the running one or to compute the mean value of
the sum of the FF's calculated using both possibilities; due to the symmetry of 
$T_H(x,y,Q^2)$ and Eq.\ (\ref{eq:5.1}) itself with respect to $x,\,y$ we will
get the same result. Of course, the second option (\ref{eq:5.4})
leaves in the NLO correction some logarithmic terms, but it leads to better
agreement with the experimental data than the choice (\ref{eq:5.3}) \cite
{ag7}. Therefore in our computations we use the option (\ref{eq:5.4}).
Since in the considering process the partonic hard subprocess contains only
one external virtual photon, we do not perform the symmetrization of the
hard-scattering amplitude. Stated differently, below we write down the
expression $[Q^2F_\pi (Q^2)]^{res}$, obtained in the framework of the
ordinary RC method.

The Borel resummed pion electromagnetic FF are determined by the formula 
\cite{ag7}  
$$
[Q^2F_\pi (Q^2)]^{res}=\frac{(16\pi f_\pi )^2}{\beta _0}\sum_{l=0}^\infty
K_lB(2+l,1) 
$$
\begin{equation}
\label{eq:5.5}
\times \sum_{n=0}^\infty K_n\int_0^\infty due^{-ut}R(u,t)B(2+n,1-u).
\end{equation}
The integrand in Eq. (\ref{eq:5.5}) has a finite number of the single IR
renormalon poles. In fact, the sums in the general expression (\ref{eq:5.5})
in practice run up to some $L_{\max },\,N_{\max }$, which for DA's with two
nonasymptotic terms are $L_{\max }=N_{\max }=4$. The maximum number of IR\
renormalon poles results from the term $\sim B(6,1-u)$. The latter can be
rewritten in the following way
$$
B(6,1-u)=\frac{\Gamma (6)\Gamma (1-u)}{\Gamma (7-u)}=\frac{120}{%
(1-u)(2-u)...(6-u)}, 
$$
making our statement evident. It is implied that the pole divergences are
removed by the principal value prescription.

In the asymptotic limit $Q^2\rightarrow \infty $ from the resummed FF we
recover the standard HSA expression. In fact, acting along the line
described in the detailed form in Sec. III, we get 
\begin{equation}
\label{eq:5.6}[Q^2F_\pi (Q^2)]^{res}\stackrel{Q^2\rightarrow \infty }{%
\longrightarrow }\frac{(16\pi f_\pi )^2}{\beta _0}B(2,1)\int_0^\infty
due^{-ut}B(2,1-u). 
\end{equation}
From Eqs. (\ref{eq:5.6}) and (\ref{eq:3.8}) we obtain 
\begin{equation}
\label{eq:5.7}[Q^2F_\pi (Q^2)]^{res}\stackrel{Q^2\rightarrow \infty }{%
\longrightarrow }\,\,\,16\pi f_\pi ^2\alpha _{{\rm s}}(Q^2), 
\end{equation}
which can be found in the context of the standard HSA by employing the pion
asymptotic DA.

To perform the numerical analysis of the scaled and resummed pion FF
$Q^2F_{\pi}(Q^2)$ and extract constraints on the pion DA's from such
consideration, we need to specify the experimental data that will be used in
the fitting procedure. Unlike the $\pi^0\gamma$ transition FF, where we have
precise CLEO data for large momentum transfers, the situation with the
$Q^2F_{\pi}(Q^2)$ is somewhat controversial. Thus the corresponding data
were obtained indirectly from the pion electroproduction experiments through
a model-dependent extrapolation to the pion pole. Moreover, the points $Q^2
>2\ {\rm GeV}^2$  are imprecise suffering from the large errors and, in addition,
there are big gaps between data points themselves. The data on the FF
$Q^2F_{\pi}(Q^2)$ reported recently by the $F_{\pi}$ Collaboration do not
change the whole picture, because the highest value of $Q^2$ at which the
measurements were performed is $Q^2=1.6\ {\rm GeV}^2$. Therefore, to improve
the precision of the $1\sigma$ analysis under the circumstances, we include
into our fitting procedure data points $Q^2 \geq 1.18\ {\rm GeV}^2$ and
slightly exceed in this way the range of validity of the RC method. But because our curves
describe the data at such low values of $Q^2$ as well,
we find our approach justified. Finally, let us note that the
datum point $Q^2= 9.77\ {\rm GeV}^2$ is also included into our
scheme and it strongly restricts  the $1\sigma$
region.

\begin{figure}
\epsfxsize=10 cm
\epsfysize=8 cm
\centerline{\epsffile{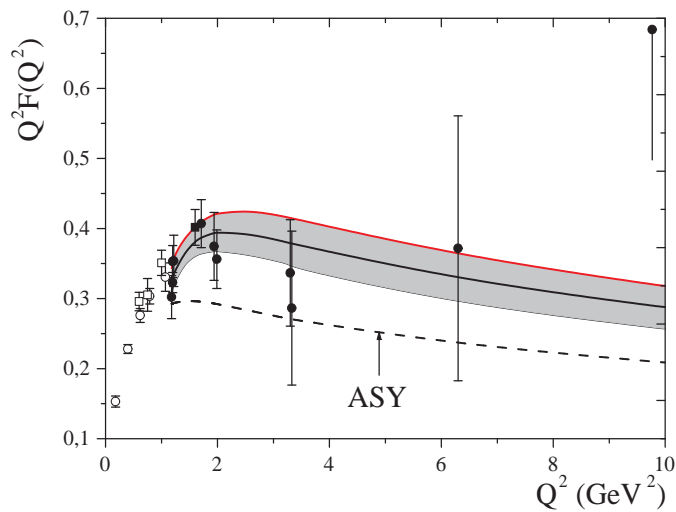}}
\caption{\label{fig:fig5}  The pion scaled and resummed electromagnetic FF
$Q^2F_\pi
(Q^2)$ as a function of $Q^2$. The shaded area is the $1\sigma$ region for
the form factor. The data are taken from Refs.
\cite{bebek} (the circles) and \cite{volmer}  (the rectangles). In the
$1\sigma$
analysis only the solid data points are used. For the central solid line the
input parameters are $b_2^0=0.23, \; b_4^0=-0.05$. For comparison the FF
obtained by means of the asymptotic DA is also plotted.}
\end{figure}

The results of our numerical calculations are
plotted in Figs.\ \ref{fig:fig5} and \ref{fig:fig2}. The $1\sigma$ area for the pion scaled
electromagnetic FF and the central curve with the Gegenbauer coefficients
$b_2^0=0.23, \, b_4^0=-0.05$ are demonstrated in Fig.\ \ref{fig:fig5}. The $1\sigma$
region for the parameters $b_2^0,\, b_4^0$ of the pion DA's
is shown in Fig.\ \ref{fig:fig2}. They obey, for example, the following
constraints

$$
b_2^0=0.16,\;\;\;b_4^0 \in [ -0.045,\ -0.05],
$$
$$
b_2^0=0.2,\;\;\;b_4^0 \in [-0.039,\ -0.058],
$$
and
$$
b_2^0=0.28,\;\;\; b_4^0 \in [-0.047,\ -0.061].
$$

The overlap of the $1 \sigma$ regions in Fig.\ \ref{fig:fig2} determines the $1 \sigma$ area in the
plane $b_4^0 - b_2^0$, within which both the $\pi^0 \gamma$
transition and the pion electromagnetic FF's are in agreement with the
corresponding data at the level of a $1 \sigma$ accuracy. As is seen, this
area is rather restricted and the values of the parameters $b_2^0$ and $b_4^0$
are

\begin{equation}
\label{eq:5.8}
b_2^0=0.235 \pm 0.035,\;\;b_4^0=-0.05 \mp 0.01.
\end{equation}

\begin{figure}
\epsfxsize=10 cm
\epsfysize=8 cm
\centerline{\epsffile{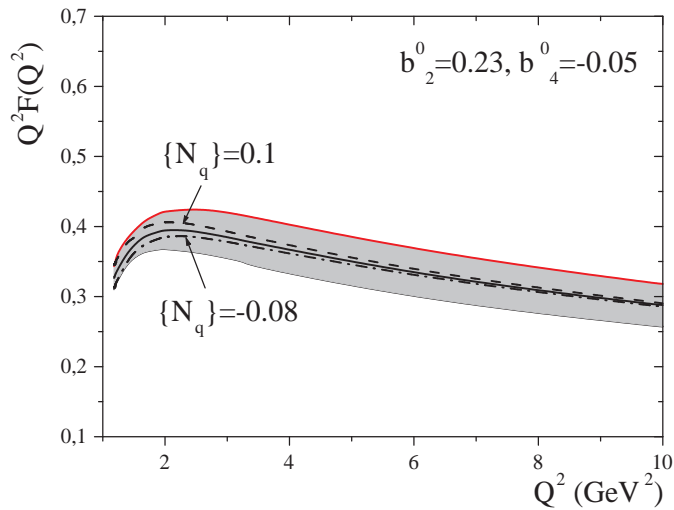}}
\caption{\label{fig:fig6}  The pion electromagnetic FF with HT ambiguities.
The solid line is the
original FF with $b_2^0=0.23,\,b_4^0=-0.05$. The broken lines include the HT
ambiguities with constants  $\{N_q\}=0.1$ (the dashed line) and
$\{N_q\}=-0.08,$ (the dot-dashed line)
$q=1,2,\ldots 6$.}
\end{figure}

The FF $Q^2F_\pi (Q^2)$ is more sensitive to HT ambiguities than the $\pi^0\gamma$ 
transition FF. Actually, in Fig.\ \ref{fig:fig6} for $b_2^0=0.23$ and 
$b_4^0=-0.05$ the scaled FF, corrected by the HT ambiguities, is plotted.
These ambiguities  for $\{N_q\}= 0.1$ and $\{ N_q \}=-0.08$ reach 
$\pm(5.6-3.6)\%$ of the FF in the region $Q^2\sim 1.2-1.6\,{\rm GeV}^2$ and 
$\pm 1 \%$ in the domain $9-10\,{\rm GeV}^2$. It is worth noting that the
estimation  of the HT ambiguities are obtained within the "$1\sigma$
constraint" (see the previous section). In
general, admissible values of the constants $\{N_q\}$ and of the input
parameters $b_2^0,\,\,b_4^0$ are strongly correlated.

\section{Concluding remarks}

\setcounter{equation}0

In this work we have calculated the power corrections to the $\pi ^0\gamma $
transition FF, originating from the end-point regions $x\rightarrow 0,1$ due
to integration of the standard HSA factorization formula with the QCD
running coupling over the longitudinal momentum fraction $x$, carrying by
the pion's quark. To this end, we have employed the RC method combined with
techniques of the IR renormalon calculus. We have used the symmetrized under
replacement $\mu _R^2\leftrightarrow \overline{\mu }_R^2$ version of the
hard-scattering amplitude of the partonic subprocess $\gamma ^{*}+\gamma
\rightarrow q+\overline{q}$.

We have obtained the Borel resummed expression $[Q^2F_{\pi \gamma
}(Q^2)]^{res}$ for the transition FF. For this purpose in the inverse Borel
transformation we have removed IR renormalon divergences by means of the
principal value prescription. Each IR renormalon pole $u_0=n$ in the
Borel transform $B[Q^2F_{\pi \gamma }](u)$ corresponds to power correction $%
\sim 1/Q^{2n}$ contained in the scaled and resummed FF. Since, in the
considering process the Borel transform has an infinite number of IR
renormalon poles, the expression (\ref{eq:2.27}), in general, contains power
corrections $\sim 1/Q^{2n},\,n=1,2,...\infty $. In numerical computations we
have truncated the corresponding series at $n_{\max }=50$. As an important
consistency check, we have proved that the result obtained within the SRC
method in the asymptotic limit $Q^2\rightarrow \infty $ reproduce the
standard HSA prediction for the transition FF. This provides justification
for the symmetrization procedure applied in the RC method.

We have compared our predictions with the CLEO data and obtained
restrictions on the input parameters of the pion DA's with one and two
nonasymptotic terms. Further constraints on the admissible set of DA's have
been extracted from the data on the pion electromagnetic FF $F_\pi (Q^2)$.
We have concluded that the pion DA's with the parameters Eq.\ (\ref{eq:5.8})
describe the experimental data on both the $Q^2F_{\pi \gamma }(Q^2)$ and $
Q^2F_\pi (Q^2)$ FF's with the $1\sigma$ accuracy.

It is important that DA's extracted from the CLEO data are suitable for
explanation of the pion electromagnetic FF, whereas in the context of the
standard HSA for describing of these FF's one has to pose on the pion DA
contradictory restrictions (or to model soft contributions to $F_{\pi}(Q^2)$ 
using mechanisms beyond the scope of the perturbative QCD). 
In fact, in the framework of the standard HSA
the pion asymptotic DA considerably underestimates the data on the
electromagnetic FF $Q^2F_\pi (Q^2)$. In order to cover a gap between the
data and theoretical curves one has to introduce model DA's with positive
and large input parameters, the Chernyak-Zhitnitsky\ DA \cite{cher} being 
one of the prominent examples.
At the same time, $\phi _{asy}(x)$ overestimates the CLEO\ data on $%
Q^2F_{\pi \gamma }(Q^2)$ and, on the contrary, model DA's with negative
input parameters are needed. The RC method solves this problem due to power
corrections taken into account in both of these quantities. Really,
the power corrections arising from the end-point integration regions at moderate 
momentum transfers significantly
enhance the pion electromagnetic FF \cite{ag4,ag7}. They also enhance
the absolute value of the NLO contribution to the FF $F_{\pi \gamma }(Q^2)$.
Since the contribution of the NLO term to $F_{\pi \gamma }(Q^2)$
is negative, power corrections effectively reduce the leading order
contribution to FF. It turns out that for some model DA's these effects lead to a
satisfactory description for both of these FF's.

The investigation performed in this work has allowed us to describe the form
factors $F_{\pi \gamma }(Q^2)$ (for $Q^2 \geq 1.64\ {\rm GeV}^2$) and 
$F_\pi (Q^2)$ (for $Q^2 \geq 1.18\ {\rm GeV}^2$) in the context of the same
theoretical scheme and by means of the same
DA's. We have achieved a quite satisfactory agreement with the available
experimental data. Theoretical computations have been carried out using the
leading order [for $F_\pi (Q^2)$] and the NLO [for $F_{\pi \gamma }(Q^2)$]
expressions for the hard-scattering amplitudes of the partonic subprocesses.
An accuracy of our theoretical predictions may be improved by including into
analyses the NLO and NNLO terms, respectively. These problems form 
directions for
improving the developed theoretical framework and require separate detailed
investigations.

\end{document}